\documentclass{aa}
\usepackage{psfig}
\usepackage{epsfig}
\def \sax {BeppoSAX}
\def \src {XB\thinspace1746-371}
\def \glob {NGC\thinspace6441}

\def \nh {N${\rm _H}$}

\def \hcm {\hbox {\ifmmode $ atom cm$^{-2}\else atom cm$^{-2}$\fi}}
\def \arcmin {\hbox{$^\prime$}}

\def\approxgt{\mathrel{\hbox{\rlap{\lower.55ex \hbox {$\sim$}}
        \kern-.3em \raise.4ex \hbox{$>$}}}}
\def\approxlt{\mathrel{\hbox{\rlap{\lower.55ex \hbox {$\sim$}}
        \kern-.3em \raise.4ex \hbox{$<$}}}}
\newcommand{\mc}{\multicolumn}
\newcommand {\Msun}{M_\odot}
\begin{document}

\thesaurus{(02.01.2; 08.09.2; 08.14.1; 10.07.3; 13.25.2; 13.25.3)}

\title{BeppoSAX spectroscopy of the globular cluster X-ray source 
\src\ (\glob)}

\author{A.N. Parmar\inst{1}
        \and T. Oosterbroek\inst{1} 
        \and M. Guainazzi\inst{1}
        \and A. Segreto\inst{2} 
        \and D. Dal Fiume\inst{3}
        \and L. Stella\inst{4}
}
\offprints{A.N.Parmar (aparmar@astro.estec. esa.nl)}

\institute{
        Astrophysics Division, Space Science Department of ESA, ESTEC,
              Postbus 299, 2200 AG Noordwijk, The Netherlands
\and
       Istituto di Fisica Cosmica ed Applicazioni all'Informatica, CNR, Via 
       U. La Malfa 153, 90146 Palermo, Italy
\and
       Istituto Tecnologie e Studio Radiazioni Extraterrestri, CNR, 
       Via Gobetti 101, 40129 Bologna, Italy
\and  
       Osservatorio Astronomico di Roma, Via Frascati 33,
       Monteporzio Catone, I-00040 Roma, Italy
}
\date{Received ; Accepted: 1999 August 18}

\maketitle

\markboth{BeppoSAX observation of \src\ (\glob)}{BeppoSAX 
observation of \src\ (\glob)}

\begin{abstract}
During a \sax\ observation of the X-ray source \src\ 
located in the globular cluster \glob\
a type I X-ray burst, parts of 4 intensity dips, and energy
dependent flaring were detected. 
The dips repeat every $5.8 \pm ^{0.3} _{0.9}$~hr 
and show no obvious energy dependence. If the dips are due to
electron scattering this energy independence implies an abundance
$>$130 times less than solar, confirming an earlier measurement.
Since the overall cluster abundance is close to solar this low
abundance is unexpected. Photoionization of the
absorbing material, obscuration of an extended source, and
variations in multiple components that combine to produce an
{\it apparent} energy independence are all excluded. Thus, the 
nature of the dips remains uncertain. 
The best-fit model to
the overall 0.3--30~keV spectrum is a disk-blackbody with a temperature of 
$\rm {2.82 \pm 0.04}$~keV, together with a 
cutoff power-law with a photon index 
of $-0.32 \pm 0.80$ and a cutoff energy of $0.90 \pm 0.26$~keV.
Absorption, consistent with the optical extinction to \glob\ of 
$\rm {(0.28 \pm 0.04)\times 10^{22}}$~atom~cm$^{-2}$ is required. 
All previous best-fit spectral models for this source are excluded at
high confidence.
The spectrum is dominated by the blackbody-like component,
with the cutoff power-law only contributing an average of 12\%
of the 1--10 keV flux. During flaring intervals the contribution
of this component decreases to $\sim$6\% with variations in
the intensity of the blackbody-like component 
being responsible for most of the flaring activity.

\end{abstract}


\section{Introduction}
\label{sect:intro}

\src\ is one of around ten low-mass X-ray binaries
(LMXRBs) known to exhibit periodic irregular dips in X-ray intensity.
These are probably caused by obscuration of the 
central X-ray
source by material located in the thickened outer region of the 
accretion disk associated with the impact of the gas stream from the
companion.
The depth, duration and spectral properties of the dips 
vary from source to source, and from cycle to cycle (see \cite{parmar88};
\cite{white95} for reviews). The study of dips provides a powerful
probe of the properties of the absorbing and emitting regions in
LMXRBs.

The spectral evolution during dips is complex. In most
sources a cold absorber model fails to adequately represent the dip
spectra which show excesses at energies $\approxlt$4~keV. 
In \src\ and X\thinspace1755-338 the dips appear highly energy independent.
Explanations for this energy independence include (1)
an abundance, or metallicity, of the absorbing material 
of at least two orders of magnitude 
less than cosmic (\cite{white84}; \cite{parmar89}). 
(2) Photoionization of the absorbing material implying that it
is located well within the outer radius of the accretion disk
(\cite{frank87}). (3) Partial covering of an extended source 
(\cite{sztajno84}). (4)
Variations in multiple continuum components which
combine to produce an {\it apparent} energy independence (\cite{church93}).
One of the components is usually taken
to be a power-law or a cutoff power-law 
and the other either a blackbody or a disk-blackbody.

A 12~hour duration uninterrupted EXOSAT observation 
of \src\ detected three apparently energy independent intensity dips
separated by $5.0 \pm 0.5$~hr (\cite{parmar89}).
The dips have a duty cycle of $\sim$20\% and are not total,
with $\sim$85\% of the 1--10~keV non-dip continuum remaining. 
Their periodic nature was confirmed by {\it Ginga} observations
which refined the dip period to be $5.73 \pm 0.15$~hr (\cite{sansom93}). 
\src\ is located within the globular cluster \glob\ whose
properties are summarized in {\cite{layden99}.
The overall cluster metallicity has been estimated
to be only a factor 3 less than cosmic (\cite{djorgovski93}). 
Variations in abundance within the
cluster are expected to be at most a factor $\sim$3 (e.g., 
\cite{norris81}). Thus, the expected abundance of the material
in \src\ differs strongly from that implied by the energy
independence of the dips of $>$150 less than solar (\cite{parmar89}).
Two X-ray bursts were detected
during the EXOSAT observation (\cite{sztajno87}).
The luminosity of these bursts is close to Eddington at the 
10.7~kpc (\cite{djorgovski93}) distance
of \glob\ indicating that the central neutron star
is directly observed. Recently, the optical counterpart to \src\
has been detected using the
Hubble Space Telescope (\cite{deutsch98}).
Its B magnitude of 18.2 implies a ratio of X-ray to optical 
luminosity, ${\rm L_x /L_{opt}}$, of $\sim$$10^3$ supporting the
view that the central source is directly observed 
(see \cite{vanparadijs95}). The counterpart exhibited $\sim$30\%
ultraviolet variability between two exposures separated by 0.5~hr.

The EXOSAT Medium-Energy 1--15~keV non-dip spectrum of \src\ can be 
represented by a power-law with a photon index, $\alpha$, of 
2.0 and low-energy absorption, N$_{{\rm H}}$, equivalent to 
$0.82 \times 10^{22}$~atom~cm$^{-2}$ (\cite{parmar89}).
\cite{callanan95} re-analyzed the EXOSAT spectrum of \src\ including
a 0.04--2.0~keV measurement from a low-energy imaging telescope. 
They find that as well as a power-law a
blackbody with a temperature, kT, of $1.08\pm ^{0.34} _{0.22}$~keV 
is required in order to obtain a satisfactory fit. 
The higher quality 2--17~keV {\it Ginga} spectrum of Sansom et al. (1993) 
cannot be successfully fit with
any of the standard models. However, the best result is also obtained with a 
power-law, but with a reduced $\chi ^2$ of 4.2 with 21 
degrees of freedom (dof).
This suggests a complex underlying spectrum, or that the effects
of energy dependent flaring observed by {\it Ginga} are significant.
In contrast, the spectrum obtained from a 
short 0.5--20~keV {\it Einstein} Solid-State 
Spectrometer and 
Monitor Proportional Counter array observation in 1979 could be well fit
by either a cutoff power-law model (${\rm E^{-\alpha}\exp-(E/E_{co})}$)
with $\alpha = 0.96 \pm 0.01$ and ${\rm E_{co} = 6.08 \pm 0.55}$~keV,
or the combination of 1.89~keV blackbody and 8.3~keV
bremsstrahlung components (\cite{christian97}).

In this {\it paper} we report on a broad-band observation
of \src\ with BeppoSAX designed to constrain the continuum shape,
investigate any spectral changes during dips, and quantify the time
variability of the source.

\section{Observations}
\label{sect:obs}

Results from the Low-Energy Concentrator Spectrometer (LECS;
0.1--10~keV; \cite{parmar97}), the Medium-Energy Concentrator
Spectrometer (MECS; 1.8--10~keV; \cite{boella97}),
the High Pressure Gas Scintillation Proportional Counter
(HPGSPC; 5--120~keV; \cite{manzo97}) and the Phoswich
Detection System (PDS; 15--300~keV; \cite{frontera97}) on-board \sax\
are presented. All these instruments are coaligned and collectively referred
to as the Narrow Field Instruments, or NFI.
The MECS consists of two grazing incidence
telescopes with imaging gas scintillation proportional counters in
their focal planes. The LECS uses an identical concentrator system as
the MECS, but utilizes an ultra-thin entrance window and
a driftless configuration to extend the low-energy response to
0.1~keV. The non-imaging HPGSPC consists of a single unit with a collimator
that remained on-source during the entire observation. The non-imaging
PDS consists of four independent units arranged in pairs each having a
separate collimator. Each collimator was alternatively
rocked on- and off-source during the observation.

The region of sky containing \src\ was observed by \sax\
on 1999 April 04 18:07 UT to April 05 20:23 UT.
Good data were selected from intervals when the elevation angle
above the Earth's limb was $>$$4^{\circ}$ and when the instrument
configurations were nominal, using the SAXDAS 2.0.0 data analysis package.
The standard PDS collimator dwell time of 96~s for each on- and
off-source position was used together with a rocking angle
of 210\arcmin.
LECS and MECS data were extracted centered on the position of \src\ 
using radii of 8\arcmin\ and 4\arcmin, respectively.
The exposures 
in the LECS, MECS, HPGSPC, and PDS instruments are 16.3~ks, 49.9~ks,
48.7~ks, and 23.3~ks, respectively. 
Background subtraction for the imaging instruments
was performed using standard files, but is not critical for such a
bright source. 
Background subtraction for the HPGSPC was 
carried out using data obtained when the instrument
was looking at the dark Earth and for the PDS using data
obtained during intervals when the collimator was offset from the 
source. 

\begin{figure*}
 \centerline{\psfig{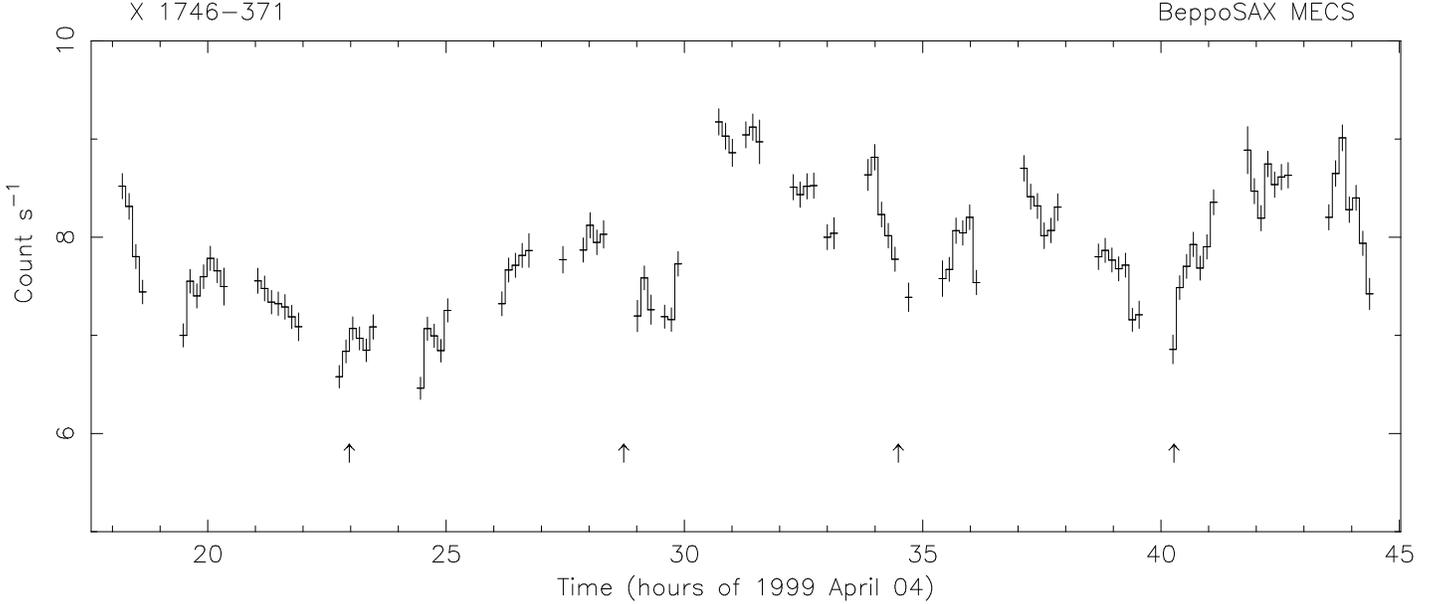}}
  \caption[]{MECS 1.8--10 keV lightcurve of \src\ with a binning
             of 512~s. The X-ray burst that occurs at around
             44~hrs is strongly truncated by the binning. The arrows
             indicate the times of minima of the folded lightcurve}
  \label{fig:lightcurve}
\end{figure*}

\section{Results}

\subsection{X-ray Lightcurve}
\label{subsect:lightcurve}

Fig.~\ref{fig:lightcurve} shows the 1.8--10~keV 
MECS lightcurve of \src\ with a binning of 512~s. Substantial 
intensity variability is present
as well as an X-ray burst, which is strongly truncated
due to the long binning. If the X-ray burst is ignored, the
average root mean square variability of the 256 ~s binned 
light curve is $7.8 \pm 0.4$\%.
The flaring activity and burst 
are not strongly evident in the 15--30~keV PDS lightcurve due
to the low count rate in this instrument.

The dipping activity seen by EXOSAT and {\it Ginga} is not readily
apparent in Fig.~\ref{fig:lightcurve}. To investigate whether dipping
is in fact present during the BeppoSAX observation
a Fast Fourier Transform (FFT) of the
1.8--10~keV MECS data excluding the bursting interval was computed.
This was rebinned in a logarithmic fashion. Fig.~\ref{fig:fft}
shows the FFT between $10^{-5}$ and $10^{-3}$~Hz. A ``spike'' is visible
at around $5 \times 10^{-5}$~Hz, together with a large excess of power
at frequencies $<$$3 \times 10^{-5}$~Hz, and a smaller feature at
$\sim$$2 \times 10^{-4}$~Hz, corresponding to the BeppoSAX orbital 
frequency. The $5 \times 10^{-5}$~Hz ``spike'' is at a frequency 
consistent with the dipping activity seen 
by EXOSAT and {\it Ginga}. This implies that similar dipping
activity is present during the BeppoSAX observation.
In order to compare this variability with that seen earlier,
a 1.8--10~keV lightcurve with a 4096~s binning was produced
and a 3rd order polynomial fit to these data. A 256~s integration 
lightcurve was then divided by the 3rd order polynomial
in order to reduce the effects of long-term variability. 
The dip
recurrence interval was then determined to be $5.8 \pm ^{0.3} _{0.9}$~hr
by epoch folding. This value is consistent with the period derived from the
{\it Ginga} observation of $5.73 \pm 0.15$~hr (\cite{sansom93}). 
The minimum of the BeppoSAX folded lightcurve occurs at 
MJD~$51273.20 \pm 0.02$. 

The variation in spectral hardness with count rate was next investigated.
The hardness ratio is defined as the count rate in the energy 
range 4.0--10.0~keV divided by that in the range 1.8--4.0~keV. 
Fig.~\ref{fig:hardness} shows the MECS hardness ratio plotted against
1.8--10~keV count rate with a binning time of 512~s.
The hardness ratio clearly
increases with increasing count rate with a probable flattening at
lower count rates. If intervals with 1.8--10~keV MECS count rate
$>$7.5~s$^{-1}$ are selected, then the hardness ratio gradient is 
$0.075 \pm 0.008$~count$^{-1}$~s (90\% confidence). 
Similarly, if intervals with count rates
below this value are selected, then the hardness ratio gradient
is $ 0.049 \pm 0.027$~count$^{-1}$~s.
Similar trends are evident in the {\it Ginga}
data reported in Sansom et al. (1993).
The low-count rate intervals tend to be associated with
dipping activity, and the small change in hardness ratio at low
count rates confirms the presence of the (almost)
energy-independent dipping discovered with EXOSAT (\cite{parmar89}).

\begin{figure}
  \centerline{\psfig{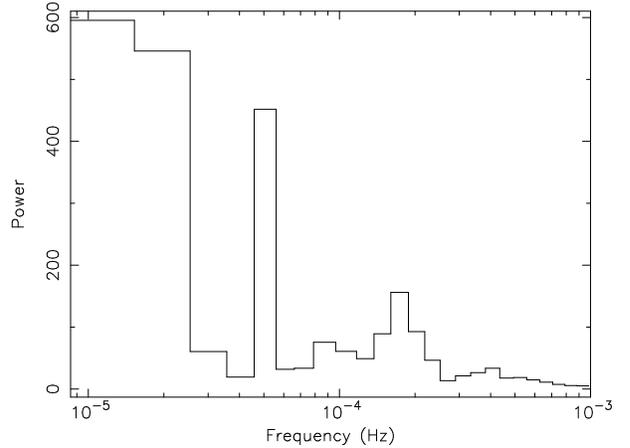}}
  \caption[]{The \src\ 1.8--10~keV MECS FFT. Power is in units of ``Leahy''
            power. The ``spike'' at around $5 \times 10^{-5}$~Hz
            corresponds to the dipping activity. The peak
            at $\sim$$2 \times 10^{-4}$~Hz is consistent with the BeppoSAX
            orbital frequency} 
  \label{fig:fft}
\end{figure}

\begin{figure}
  \centerline{\psfig{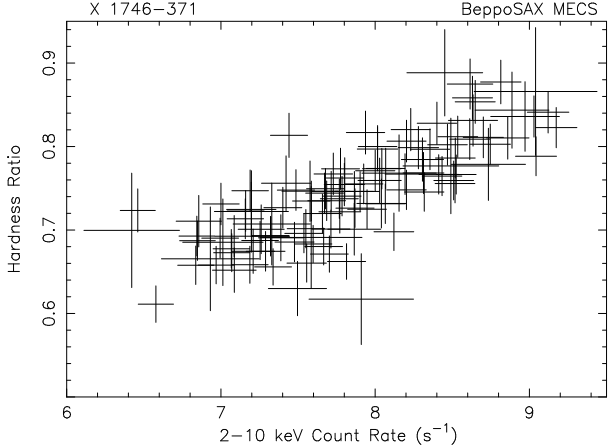}}
  \caption[]{The \src\ MECS hardness ratio plotted against 1.8--10~keV count
             rate. The hardness ratio increases with increasing count rate
             and shows a tendency to flatten at the lowest count rates}
  \label{fig:hardness}
\end{figure}

Fig.~\ref{fig:folded} shows the 1.8--10~keV MECS lightcurve folded on the 
best-fit dip period given above. 
The interval corresponding to the X-ray burst has
been excluded and the data divided by the best-fit 3rd-order polynomial. 
The peak-to-peak
variability of the folded lightcurve is $14 \pm 1$\%. 
With the exception of the large
peak at phase $\sim$0.3, the BeppoSAX folded lightcurve is
similar to those of Sansom et al. (1993). This similarity provides
further support for
the presence of dips during the BeppoSAX observation.
Examination of Fig.~\ref{fig:lightcurve} reveals
that the high count rate at $\sim$31~hrs contributes significantly to
the peak in the folded lightcurve, but that similar intensity maxima
are not seen during the other 4 phase 0.3 intervals observed. This
suggests that the peak at phase $\sim$0.3 in the folded lightcurve 
results from intrinsic
variability, rather than an orbital modulation. Given the sharpness
of the lightcurve and hardness ratio minima, Fig.~\ref{fig:folded}
suggests that dipping activity probably occurs over $\sim$15\%
of the orbital cycle.

\begin{figure}
  \centerline{\psfig{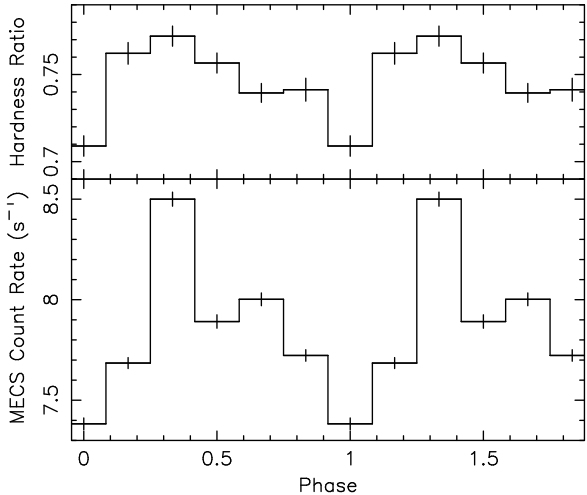}}
  \caption[]{The \src\ 1.8--10~keV MECS lightcurve (lower panel) and hardness
             ratio (upper panel) folded over the
             5.8~hr dip period, after removal of the long-term 
             variability and normalized to the mean count rate. 
             The interval corresponding to the X-ray 
             burst is excluded}
  \label{fig:folded}
\end{figure}

\subsection{Overall Spectrum}
\label{subsect:spectrum}

The overall spectrum of \src\ was first investigated by simultaneously
fitting data from all the \sax\ NFI. The interval corresponding to
the burst was excluded.
The LECS and MECS spectra were rebinned to oversample the full
width half maximum of the energy resolution by
a factor 3 and to have additionally a minimum of 20 counts 
per bin to allow use of the $\chi^2$ statistic. 
The HPGSPC and PDS 
spectra were rebinned using the standard techniques in SAXDAS.
Data were selected in the energy ranges
0.3--4.0~keV (LECS), 1.8--10~keV (MECS), 8.0--20~keV (HPGSPC),
and 15--30~keV (PDS) 
where the instrument responses are well determined and sufficient
counts obtained. 
This gives
background-subtracted count rates of 2.9, 7.8, 3.3 and 0.8~s$^{-1}$ 
for the LECS, MECS, HPGSPC, and PDS, respectively.
The photoelectric absorption
cross sections of \cite{morisson83} and the
solar abundances of \cite{anders89} are used throughout

\begin{table*}
\caption[]{\src\ overall spectral fit results.
\nh\ is in units of $\rm {10^{22}}$ atom $\rm {cm^{-2}}$.
${\rm r_{in}({\cos}i)^{0.5}}$ is in units of km for a distance
of 10.7~kpc. 90\% confidence limits are given. PL = power-law,
DBB = disk-blackbody}
\begin{flushleft}
\begin{tabular}{lccccccr}
\hline\noalign{\smallskip}
Model & \hfil N$_{\rm {H}}$ \hfil & kT (keV) & kT${\rm _{BB}}$ (keV) &$\alpha$
& $\rm{E_{co}}$ (keV) & ${\rm r_{in}({\cos}i)^{0.5}}$ & $\chi^2$/dof \\
\noalign{\smallskip\hrule\smallskip}
Bremss + blackbody & $0.44 \pm 0.015$ & $6.39 \pm 0.20$ &
$1.85 \pm 0.04$ & \dots & \dots & \dots & 180.3/108 \\
PL + DBB & $0.53 \pm 0.03$ & \dots & $2.68 \pm
0.03$ & $ 2.59 \pm 0.09 $ & \dots & $1.15 \pm 0.05$ & 179.3/108 \\
Cutoff PL + blackbody & $0.21 \pm 0.02$ & \dots & $0.50 \pm 0.02$ &
$0.18 \pm 0.04$ & $3.5 \pm 0.7 $ & \dots & 112.3/107\\ 
Cutoff PL + DBB & $0.28 \pm 0.04$ & \dots & $2.82 \pm
0.04$ & $ \llap{$-$}0.32 \pm 0.80 $ & $0.90 \pm 0.26$ & $1.03 \pm 0.11$ 
& 108.0/107 \\
\noalign{\smallskip\hrule\smallskip}
\end{tabular}
\end{flushleft}
\label{tab:spec_paras}
\end{table*}

Initially, simple models were tried, including absorbed power-law,
thermal bremsstrahlung and cutoff power-law 
models. 
Factors were included in the spectral fitting to allow for normalization 
uncertainties between the instruments. These factors were constrained
to be within their usual ranges during the fitting. 
A power-law with $\alpha = 2.2$ and a 
8.6~keV bremsstrahlung  
give unacceptable fits with $\chi ^2$s of 16000 and 3000 for 110 dof, 
respectively.
A cutoff power-law model with $\alpha = 0.74$ and ${\rm E_{co} = 4.9}$~keV
gives a better fit with a $\chi ^2$ of
370 for 109 dof. Next more complex models consisting of a 
bremsstrahlung and a blackbody and a power-law and a disk-blackbody 
(\cite{mitsuda84}; \cite{makishima86})
were tried for comparison with previous results.
The disk-blackbody model assumes that the gravitational energy
released by the accreting material is locally dissipated into
blackbody radiation, that the accretion flow is continuous throughout
the disk, and that the effects of electron scattering are negligible.
There are only two parameters of the model, 
${\rm r_{in}({\cos}i)^{0.5}}$ where ${\rm r_{in}}$ is the innermost
radius of the disk, i is the inclination angle of the disk and
${\rm kT_{BB}}$ the blackbody effective temperature at ${\rm r_{in}}$. 
Both these models gave somewhat better fits with $\chi ^2$s
of 180.3 and 179.3 for 108 dof. 

Inspection of the residuals of the power-law and disk-blackbody fit
suggests that the power-law should be replaced by a cutoff power-law.
This results in an acceptable $\chi ^2$ of 108.0 for 107 dof.
If the disk-blackbody is replaced by a 0.50~keV blackbody, the 
fit is still acceptable with a $\chi ^2$ of 112.3 for 107 dof.
An F-test indicates that the probability of chance improvement is 87\%.
The blackbody radius is $12.5 \pm 0.3$~km and 
${\rm r_{in}({\cos}i)^{0.5}}$ from the disk-blackbody fit 
is $1.03 \pm 0.11$~km, both for a distance of 10.7~kpc.
Since i is likely to be $\approx$70${\rm ^\circ}$ (see 
Sect.~\ref{sect:discussion}) the disk-blackbody fit implies an unusually small
emitting region of $\approx$2~km radius.
The fit results are summarized in Table~\ref{tab:spec_paras} and the
fit using the cutoff power-law and disk-blackbody model is shown in 
Fig.~\ref{fig:spectrum}. 
The overall 1--10~keV flux is $8.6 \times 10^{-10}$~erg~cm$^{-2}$~s$^{-1}$,
which corresponds to a luminosity of $1.2 \times 10^{37}$~erg~s$^{-1}$ at
a distance of 10.7~kpc. The cutoff power-law
is much weaker than the disk-blackbody 
(or blackbody), only contributing 12\% (11\%) of the
1--10~keV luminosity, and with a maximum contribution around 1~keV
(see Fig.~\ref{fig:spectrum}). Since this component is so weak its
overall shape is poorly constrained and if it
is replaced by e.g. a 3.5~keV thermal bremsstrahlung the resulting
$\chi ^2$ is 126.4 for 108~dof. Similarly, replacing it by a 0.55~keV blackbody
gives a $\chi ^2$ of 114.9 for 108 dof, marginally worse than the
cutoff power-law and blackbody fit. 
However, replacing the faint component with the {\sc comptt} 
model in {\sc xspec}, used to describe the Comptonization of cool photons
in a hot plasma, does not give an acceptable fit.
The 90\% confidence upper limit to the equivalent width of
a narrow Fe line at 6.5~keV is 15~eV. Absorption equivalent to
$(0.28 \pm 0.04) \times 10^{22}$~atom~cm$^{-2}$ is required. The optical
extinction to \glob, ${\rm A_v}$, is 1.30 magnitudes (\cite{djorgovski93})
which corresponds to an ${\rm N_H}$ of $0.23 \times 10^{22}$~atom~cm$^{-2}$
using the relation in \cite{predehl95}. Thus, there is no strong evidence
for significant absorption intrinsic to the X-ray source.

\begin{figure*}
  \centerline{
  \hbox{
   \psfig{figure=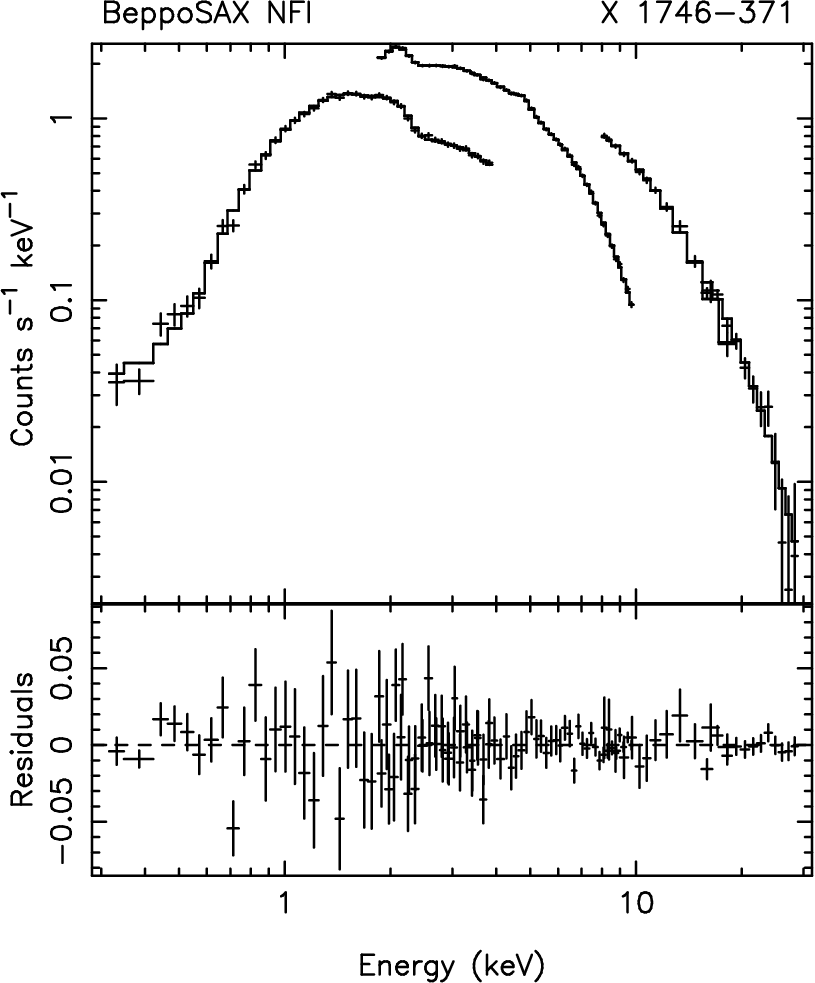,width=9.0cm,angle=-90}
   \hspace{1.0cm}
   \psfig{figure=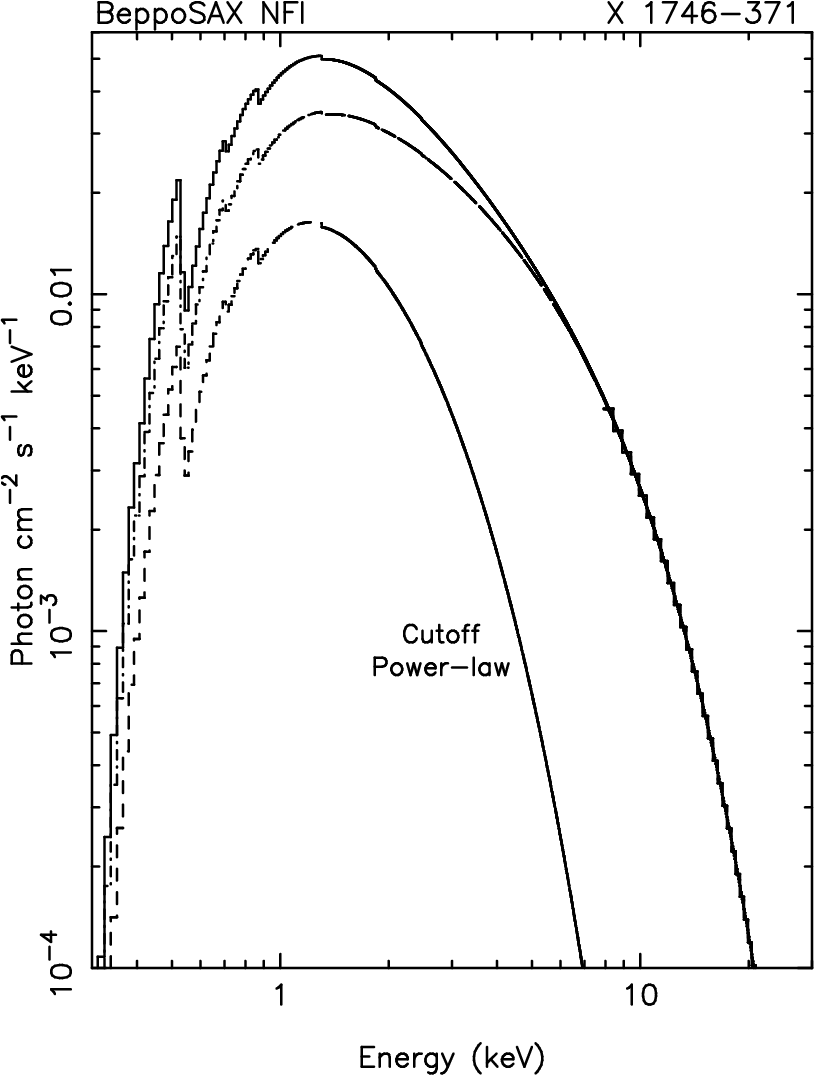,width=9.0cm,angle=-90}}}
  \caption[]{The overall \src\ NFI spectrum together with the best-fit 
             cutoff power-law and disk-blackbody model fit 
             (see Table~\ref{tab:spec_paras}). 
             The lower-left panel shows the 
             fit residuals in units of counts~s$^{-1}$~keV$^{-1}$.
             The right panel shows the assumed photon spectrum with
             the contributions of the cutoff power-law and disk-blackbody
             components indicated separately}
  \label{fig:spectrum}
\end{figure*}

\subsection{Intensity Selected Spectra}
\label{subsect:intensity_spectrum}

Fig.~\ref{fig:hardness} illustrates that the hardness ratio depends on
intensity. In order to investigate the
energy dependence of this variability, a series of intensity selected 
spectra were produced.
Intervals corresponding to MECS 1.8--10~keV
count rates of $<$7.2, 7.2--7.6, 7.6--8.1, 8.1--8.6, $>$8.6~s$^{-1}$, 
before any long-term trend removal when
the data are accumulated with a binning of 256~s, were determined.
These intervals were used to extract a set of corresponding 5 NFI 
spectra which
were then rebinned and energy selected in the same way as in 
Sect.~\ref{subsect:spectrum}.
Since the cutoff power-law and disk-blackbody 
model provides the best fit to the overall
spectrum, this model was also fit to these new spectra.

\begin{figure}
  \centerline{\psfig{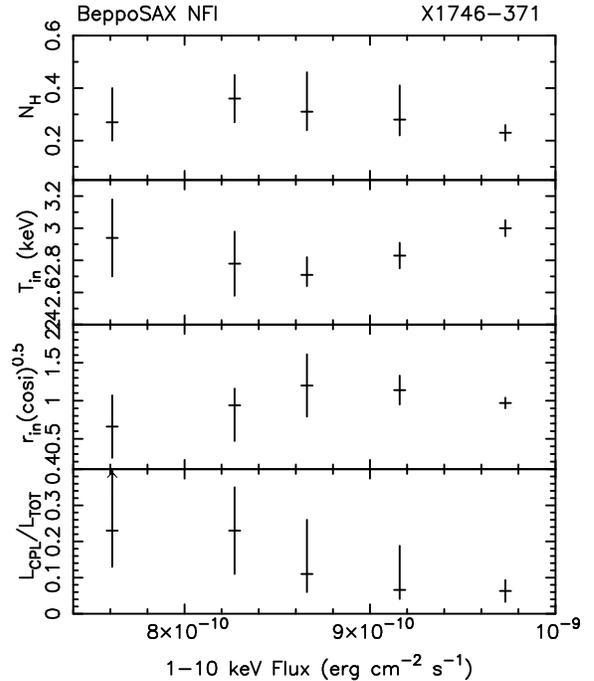}}
  \caption[]{The dependence of ${\rm N_H}$, ${\rm T_{in}}$, 
             ${\rm r_{in}(\cos i)^{0.5}}$ and the ratio of 1--10~keV
             fluxes in the cutoff power-law and disk-blackbody components
             against intensity. The units are the same as in 
             Table~\ref{tab:intensity_paras}}
         \label{fig:intensity}
\end{figure}

The fit results are given in Table~\ref{tab:intensity_paras}. With the
exception of the lowest intensity selection, the spectra are well
fit by the model. Examination of the residuals of the
lowest intensity fit shows that there are no obvious systematic
deviations between model and data, and that the data just appear to 
be ``noisy''. Allowing different amounts of low-energy absorption
for the two components does not produce a significant improvement
in fit quality. If the minimum absorption of the cutoff power-law
is constrained to be equal to the intersteller value of
$0.23 \times 10^{22}$~atom~cm$^{-2}$ (see Sect.~\ref{subsect:spectrum}),
the 90\% confidence limit to any additional absorption on the
disk-blackbody is $0.48 \times 10^{22}$~atom~cm$^{-2}$.
The best-fit values of ${\rm N_H}$, ${\rm T_{in}}$, 
${\rm r_{in}(\cos i)^{0.5}}$ and the ratio of 1--10~keV
fluxes in the cutoff power-law compared to the total
are shown plotted against total 1--10~keV flux in Fig.~\ref{fig:intensity}.
There is no systematic change in ${\rm N_H}$ with
intensity. The values of ${\rm T_{in}}$ and ${\rm r_{in}(\cos i)^{0.5}}$ 
are strongly correlated and have no obvious
luminosity dependence.
The spectral parameters of the cutoff power-law are not well determined
and are also strongly correlated, and so we consider only
the intensity dependence of the luminosity of this component. 
Table~\ref{tab:intensity_paras} and Fig.~\ref{fig:intensity} show
that the contribution of the cutoff power-law to the total decreases
from $\sim$23\% at a flux of 
$7.6 \times 10^{-10}$~erg~cm$^{-2}$~s$^{-1}$~keV$^{-1}$ to $\sim$6\%
at a flux of $9.7 \times 10^{-10}$~erg~cm$^{-2}$~s$^{-1}$~keV$^{-1}$.
This means that the intensity of the cutoff power-law decreased by a 
factor $\sim$3 as the overall intensity increased by 28\% and that
the majority of the flaring activity may be modeled by changes in
the intensity of the disk-blackbody. A similar result is
obtained if the disk-blackbody is substituted by a blackbody.

\begin{table*}
\caption[]{\src\ intensity selected NFI spectral fit results using
the disk-blackbody and cutoff power-law model.
\nh\ is in units of $\rm {10^{22}}$ atom $\rm {cm^{-2}}$.
${\rm r_{in}({\cos}i)^{0.5}}$ is in units of km for a distance
of 10.7~kpc. ${\rm L_{CPL}/ L _{TOT}}$ is the 1--10~keV ratio of the
cutoff power-law to the total luminosity. 90\% confidence limits are 
given}
\begin{flushleft}
\begin{tabular}{lccccc}
\hline\noalign{\smallskip}
Parameter & \mc{5}{c}{MECS Count Rate (s$^{-1}$)} \\
 & $<$7.2 & 7.2--7.6 & 7.6--8.1 & 8.1--8.6 & $>$8.6 \\
\noalign{\smallskip\hrule\smallskip}
MECS Exp. (ks) & 7.9 & 10.0 & 14.0 & 10.2 & 7.0 \\ 
${\rm N_H}$  & $0.27 \pm ^{0.13} _{0.07}$ & 
$0.36 \pm 0.09 $ & $0.31 \pm ^{0.15} _{0.07}$ & $0.28 \pm ^{0.13} _{0.06}$ &
$0.23 \pm 0.03$ \\
$\alpha $ & $0.25 \pm 1.20$ & $0.70 \pm ^{1.0} _{1.5}$ & $0.0 \pm 2.0$ &
$\llap{$-$}0.40 \pm 2.5$ & $\llap{$-$}3.0 \pm ^{1.6} _{1.0} $ \\
${\rm E_{co}}$ (keV) & $1.4 \pm 0.7 $ & $1.7 \pm ^{6.0} _{1.0}$ &
$1.0 \pm ^{4.5} _{0.5}$ & $0.8 \pm ^{2.7} _{0.4}$ & $0.41 \pm 0.03$ \\ 
${\rm T_{in}}$ (keV) & $2.94 \pm 0.24$ & $ 2.78 \pm 0.20$ & 
$ 2.71 \pm ^{0.11} _{0.07}$ & $2.83 \pm 0.08$ & $3.00 \pm 0.05$ \\
${\rm r_{in}({\cos}i)^{0.5}}$ & $0.66 \pm 0.41$ & $0.94 \pm ^{0.22} _{0.47}$ &
$1.20 \pm 0.41$ & $1.14 \pm 0.19$ & $ 0.97 \pm 0.07$\\
${\rm L_{CPL}/ L _{TOT}}$ & $0.23 \pm ^{0.51} _{0.10}$ &
$0.23 \pm 0.12 $ & $0.11 \pm ^{0.15} _{0.05}$ & $ 0.066 \pm ^{0.122} _{0.025} $
& $0.063 \pm 0.003$\\
$\chi ^2$/dof & 122.3/103 & 100.2/103 & 109.1/104 & 94.3/104 & 106.7/104 \\
\noalign{\smallskip\hrule\smallskip}
\end{tabular}
\end{flushleft}
\label{tab:intensity_paras}
\end{table*}

\subsection{Dip Spectrum}
\label{subsect:dip_spectrum}

In order to investigate the spectral behavior during dips an NFI spectrum
corresponding to the lowest bin of the folded lightcurve 
(Fig.~\ref{fig:folded}) was extracted. This covers phases
0.92--1.08 of the ephemeris given in Sect.~\ref{subsect:lightcurve}.
This selection was chosen since both the folded lightcurve and hardness
ratios (Fig.~\ref{fig:folded}; upper panel) show large differences between
this and adjacent bins. This suggests that the dipping activity
occurs primarily within this single phase bin.
The spectrum was rebinned and energy selected in the same way as in 
Sect.~\ref{subsect:spectrum}.
The resulting exposure times are 2.3~ks, 8.2~ks, 7.6~ks, 3.7~ks for
the LECS, MECS, HPGSPC, and PDS, respectively.

Since dips are believed to be due to obscuration of the central 
X-ray source by intervening material, the usual
method of analyzing the spectral changes during dipping activity is to
fix the spectral shape to that outside of dips and only allow 
the amount of absorption and/or scattering to vary 
(e.g., \cite{parmar86}; \cite{church98}). In the case of \src, this
procedure is likely to give incorrect results 
since the non-dip spectral shape depends
on luminosity (see Fig.~\ref{fig:hardness})
and it is not possible to easily separate intrinsic variability
from dipping activity. Instead, the
cutoff power-law and disk-blackbody model
was fit to the NFI dip spectrum with all the parameters free, as before.
The resulting $\chi ^2$ is 117.5 for 102~dof.
Unsurprisingly, the best-fit parameter values are consistent with those 
obtained for the lowest ($<$7.5 count~s$^{-1}$) intensity selected
interval. The best-fit value of ${\rm N_H}$ is $(0.25 \pm ^{0.09} _{0.05})
\times 10^{22}$~atom~cm$^{-2}$ which implies that any change in
${\rm N_H}$ during the dips is $\approxlt$$0.06 \times 10^{22}$~atom~cm$^{-2}$,
when compared to the overall value. For comparison, the column
of electrons required to produce the observed mean reduction in
flux during the dips of 5\% by scattering is $8 \times 10^{22}$~cm$^{-2}$.
This difference implies an abundance anomaly of $>$130 less
than solar, comparable with the EXOSAT value of $>$150 less
than solar (\cite{parmar89}), which was obtained with the spectral 
parameters held fixed.

\section{Discussion}
\label{sect:discussion}

We have observed \src\ in the 0.3--30~keV energy range using \sax. 
Although the dipping activity observed by EXOSAT and {\it Ginga} 
is not readily apparent in the BeppoSAX lightcurve, an FFT of the 1.8--10~keV
MECS data reveals evidence for excess variability on a timescale of 
$\sim$5~hr. A folding analysis gives a best period of 
$5.8 \pm ^{0.3}_{0.9}$~hr, consistent with earlier EXOSAT and {\it Ginga}
measurements. When the effects of intrinsic variability are ignored, the
folded lightcurve is similar to those observed by {\it Ginga}. The above
similarities imply that dipping is present during the BeppoSAX observation.
Examination of the EXOSAT and {\it Ginga} lightcurves reveals
that it is unsurprising that the dips are so hard to see in the BeppoSAX
data. This is because the {\it Ginga} mean count rate is $\sim$190~s$^{-1}$, 
compared to 7.8~count~s$^{-1}$ for the MECS, while the continuous 
observation of EXOSAT confers an obvious advantage.

The 0.3--30~keV \src\ spectrum is clearly complex and requires at least
two continuum components. 
The best-fit model for the overall spectrum 
consists of a disk-blackbody together with a
cutoff power-law. However, we cannot exclude that the soft
component is actually a blackbody, rather than a disk-blackbody, or
that the cutoff power-law actually has another form. 
A better resolution of this spectral complexity awaits further studies.
In contrast to e.g., the dip source X\thinspace1916-053 which was detected
to energies of $\approxgt$100~keV by BeppoSAX and where
the thermal component contributes 20\% of the 1--10~keV flux (\cite{church98}),
the \src\ spectrum is dominated by the thermal component
resulting in the source being only detected to $\sim$30~keV.
The source becomes harder during flaring intervals, confirming the {\it Ginga}
result of Sansom et al. (1993). Changes in the disk-blackbody component
are responsible for most of the flaring activity. During flares the
intensity of the cutoff power-law component decreases. 
 
We can exclude at high confidence the best-fit models obtained using EXOSAT 
(PL, or cutoff PL; \cite{parmar89}, PL + blackbody; Callanan et al. 1995), 
{\it Ginga}
(PL; \cite{sansom93}) and {\it Einstein} (cutoff PL, or bremsstrahlung
and blackbody; \cite{christian97}). However, it is possible that
we are observing \src\ in an unusual state and at other
times the spectrum could be represented by a more simple model.  
Some support for this idea comes from the HEAO-1 A4 catalog where \src\
is detected in the 40--80~keV energy range, but
with a similar overall X-ray level as during the BeppoSAX observation
(\cite{levine84}). In 
a survey of the hard X-ray properties of LMXRBs using results from the 
A4 catalog, \cite{vanparadijs94} show that the high-energy spectra 
of LMXRB become
softer with increasing luminosity, and that the properties of \src\ appear 
typical of systems with similar luminosity. It has been noted that
at luminosities $10^{36}$--$10^{37}$~erg~s$^{-1}$ the spectra
of burst sources evolve from hard power-law (-like) to 
softer exponential (-like) shapes (\cite{white88}; \cite{barret94}). 

In general, the spectra of LMXRB
are interpreted in terms of an accretion disk around
a weakly magnetized neutron star. This gives rise to two spectral
components (e.g., \cite{white88}; \cite{barret94}). A ``thermal-like''
component represents the emission from an optically thick region
such as the neutron star surface (\cite{mitsuda84}) or
a boundary layer between the accretion disk and the neutron star
surface (\cite{white88}).
The second component is often represented by a cutoff power-law which is 
an approximation of the spectral shape expected from the
Comptonization of cool photons on hot electrons. Based on the 
spectral changes during dips, where the blackbody-like component
is obscured more rapidly and deeply than the Comptonized component
(e.g., \cite{church98}), and the 
discovery of hard X-ray time lags (e.g., \cite{ford99}),
this component is almost certainly extended in a number of sources. 
There are at least two possible explanations as to why the Comptonized
component is so weak in \src. (1) It is possible that the
luminosity of \src\ is close to the ``critical'' value where the 
spectra of LMXRBs may change from a power-law (-like) to an exponential (-like)
form. If this is so, a small change in the accretion rate may produce
a large change in spectral shape. (2) The geometry of \src\ may
be somehow different from the majority of LMXRBs. In particular,
due to the inclination angle through which the system is viewed (see
below) it may be possible that a significant fraction of an extended
Comptonizing region is hidden behind the accretion disk.
This could occur if e.g. a relatively small Comptonizing region 
surrounds the neutron star and is partly obscured by the flared inner
regions of the accretion disk.

The orbital parameters of \src\ can be estimated assuming that the
system contains a compact object of mass ${\rm 1.4\Msun}$, that the orbital
period is 5.73~hr and that the companion is a low-mass zero-age 
main-sequence star filling its Roche lobe. In this case the relation
between orbital period in hours, ${\rm P_{hr}}$, and 
companion mass, ${\rm M_c}$, is ${\rm M_c \simeq 0.11 \, P_{hr}}$.
For the \src\ system this
implies ${\rm M_c \simeq 0.65 \, \Msun}$. The separation between the
X-ray source and its companion is $\sim$$1.5 \times 10^{11}$~cm.
If the accretion disk fills 70\% of its Roche lobe then its radius
is $\sim$$7 \times 10^{10}$~cm.
The absence of X-ray eclipses implies that the line of sight
is inclined by $\approxgt$20${\rm ^\circ}$ from the orbital plane. 
Similarly, the dipping behavior requires that there is structure
in the accretion disk located $\approxgt$20${\rm ^\circ}$ from the orbital
plane.

The dipping behavior observed from \src\ is complex and poorly
understood.
During dips the change in low-energy absorption is 
$\approxlt$$0.06 \times 10^{22}$~atom~cm$^{-2}$ which implies an
abundance of $>$130 less than solar, consistent with 
the previous EXOSAT estimate (\cite{parmar89}) and the non-detection
of an Fe line in the X-ray spectrum. 
However, since the overall cluster metallicity is
only a factor 3 less than solar (\cite{djorgovski93}),
it is difficult to understand how the metallicity of the
absorbing material in \src\ could be so low. The other
mechanisms that could produce energy independent dips discussed in
Sect.~\ref{sect:intro} can also be excluded.
Photoionization of the elements responsible for
photoelectric absorption in the X-ray band is unlikely to be a viable
mechanism unless the material is located very close to the neutron star. 
Following arguments presented in \cite{mason85} for
X\thinspace1755-338, only material within $\sim$$10^{10}$~cm of the
central X-ray source in \src\ is likely to fully ionized, whereas the
radius of the accretion disk is expected to be 
$\sim$$7 \times 10^{10}$~cm. Partial obscuration of an extended
source is unlikely to be applicable given the high ${\rm L_x / L_{opt}}$
ratio of the system which implies that the neutron star is directly
observed.  

In the case of X\thinspace1755-338, where the energy independence of the
dips implies an abundance $>$600 times less than solar (\cite{white84}),
a two component model consisting of a blackbody and a power-law 
which combine in such a way to produce the {\it apparent} energy
independence, although the dips are caused (primarily) by absorption
of the blackbody component, has been successfully applied (\cite{church93}).
Different amounts of absorption might be expected from a 
point-like source located
close to the center of the accretion disk and an extended (Comptonized)
region. 
When the two component model is fit to the dip spectrum there is no
evidence for significant extra absorption of the disk-blackbody 
component. This implies that the multiple component model used to
describe the X\thinspace1755-338 dips in terms of absorption by
material with solar abundance is not applicable here. Thus the nature
of the dips and the reason for their apparent energy independence
remains uncertain.

\begin{acknowledgements}
The \sax\ satellite is a joint Italian-Dutch programme. 
We thank the staffs of the \sax\ Science Data and
Operations Control Centers for help with these observations. 
M.~Guainazzi acknowledges an ESA Fellowship. 
\end{acknowledgements}

\end{document}